%% file: main.tex
\documentclass[conference]{IEEEtran}

\usepackage{cite}
\usepackage{amsmath,amssymb,amsfonts}
\usepackage{algorithmic}
\usepackage{graphicx}
\usepackage{textcomp}
\usepackage{xcolor}
\usepackage{algorithm}
\usepackage{algorithmic}
\usepackage{pgfplots}
\usepackage{pgfplotstable}
\pgfplotsset{compat=1.18}
\usepackage{pgfplots}
\pgfplotsset{compat=1.18}
\pgfplotsset{table/search path={.}}
\usepackage{float}

\usepackage{amsthm}
\newtheorem{theorem}{Theorem}

\usepackage{tikz}
\usetikzlibrary{calc}

\usepackage{pgfplots}
\pgfplotsset{compat=1.18}

\usepackage{booktabs}

\graphicspath{{paper/images/}}

\newcommand{\prot}{{\sc Coalesce}}

\usepackage[hidelinks]{hyperref}

\def\BibTeX{{\rm B\kern-.05em{\sc i\kern-.025em b}\kern-.08em
    T\kern-.1667em\lower.7ex\hbox{E}\kern-.125emX}}

\begin{document}

\title{\prot\ : Hypergraph-Based Multi-Party Payment Channel}

\author{
\IEEEauthorblockN{Ayush Nainwal}
\IEEEauthorblockA{
IIT Jodhpur\\
Jodhpur, India\\
iayush.n2@gmail.com
}
\and
\IEEEauthorblockN{Atharva Kamble}
\IEEEauthorblockA{
IIT Jodhpur\\
Jodhpur, India\\
atharvakamble028@gmail.com
}
\and
\IEEEauthorblockN{Nitin Awathare}
\IEEEauthorblockA{
IIT Jodhpur\\
Jodhpur, India\\
nitina@iitj.ac.in
}
}


\maketitle

\begin{abstract}
Public blockchains inherently offer low throughput and high latency, motivating off-chain scalability solutions such as Payment Channel Networks (PCNs). However, existing PCNs suffer from \emph{liquidity fragmentation}—funds locked in one channel cannot be reused elsewhere—and \emph{channel depletion}, both of which limit routing efficiency and reduce transaction success rates. Multi-party channel (MPC) constructions mitigate these issues, but they typically rely on leaders or coordinators, creating single points of failure and providing only limited flexibility for inter-channel payments.

We introduce \textbf{Hypergraph-based Multi-Party Payment Channels (\prot)}, a new off-chain construction that replaces bilateral channels with collectively funded hyperedges. These hyperedges enable fully concurrent, leaderless intra- and inter-hyperedge payments through verifiable, proposer-ordered DAG updates, offering significantly greater flexibility and concurrency than prior designs. Hence our, design eliminates routing dependencies, avoids directional liquidity lock-up, and does not require central monitoring services such as watchtowers.

Our implementation on a \(150\)-node intra-hyperedge achieves a transaction success rate of approximately \(94\%\) under heavy load (larger payment sizes), while full hyperedge evaluation over a \(15{,}000\)-node network sustains success rates in the range of \(85\%\) to \(95\%\), without HTLC expiry or routing failures, highlighting the robustness of \prot.
\end{abstract}

\begin{IEEEkeywords}
payment channels, blockchain, hypergraphs, scalability, distributed systems
\end{IEEEkeywords}

\input{paper/sections/introduction}

\input{paper/sections/background_and_model}

\input{paper/sections/intra_hyperedge}
\input{paper/sections/inter_hyperedge}
\input{paper/sections/settlement_and_balance_update}
\input{paper/sections/channel_closure}
\input{paper/sections/security_analysis}
\input{paper/sections/evaluation}
\input{paper/sections/related_work}
\input{paper/sections/future_work}
\input{paper/sections/conclusion}

\bibliographystyle{IEEEtran}
\bibliography{references}

\end{document}

%% file: paper/sections/introduction.tex
\section{Introduction}

Public blockchains such as Bitcoin and Ethereum enable decentralized value transfer, but their throughput and latency are limited by on-chain consensus and block-size constraints \cite{nakamoto2008bitcoin,buterin2014ethereum}. Payment Channel Networks (PCNs) move frequent payments off-chain into pre-funded channels and route value over paths of bilateral channels, significantly reducing on-chain interactions \cite{PCN2,PCN3, 9912301}. However, PCNs suffer from a practical limitation we call \emph{liquidity fragmentation}: funds are tied to individual pairwise channels and cannot be reused across a user's other channels without explicit rebalancing. Over time this leads many channels to become effectively one-way and reduces end-to-end transaction success rates in practice \cite{Rebal1,Revive}.


Prior work addresses this problem in three broad ways. First, virtual channels and related on-chain constructions reduce per-payment on-chain cost at the expense of extra setup or protocol complexity \cite{9519487,Sprites2017,perun2019}. Second, routing and congestion-aware forwarding attempt to avoid depleted links at routing time \cite{SpiderRouting,Flash,Flare}. Third, active rebalancing protocols move liquidity between channels proactively to improve future success rates \cite{Revive,Rebal1,Rebal2:HubsAndRebalancing, rebal3}. These techniques improve performance in many settings, but they do not change the fundamental fact that liquidity is allocated at the granularity of pairwise channels: rebalancing itself consumes resources and often requires additional transactions that do not correspond to immediate user payments.

Multi-party channels (MPCs) generalize pairwise channels in a conventional PCN by funding a group of users with a single collective UTXO, allowing intra-group transfers without a fresh on-chain transaction \cite{perun2019,10024888,DBLP:journals/iacr/AumayrAM22}. MPCs can improve liquidity utilization, but existing designs commonly rely on coordinators or leader-based routines, which introduce single points of failure and restrict concurrent updates across participants \cite{perun2019,chanfactory2018}. In practice, ensuring atomic multi-hop transfers across independently funded groups typically requires either an on-chain coordination step or a trusted coordinator to impose a single ordering over cross-group transaction execution, which limits scalability and increases latency \cite{Sprites2017,perun2019}.



We present \prot, which replaces bilateral payment channels in conventional PCNs with multi-party payment channels. The resulting network forms a hypergraph, where each MPC in \prot\ corresponds to a hyperedge.\footnote{We use the terms MPC (or simple payment channel) and hyperedge interchangeably throughout the paper.}



\prot\ addresses a practical limitation of existing payment-channel designs: moving value across groups often requires explicit coordination, serialization, or trusted infrastructure, which undermines decentralization and increases latency. Prior multi-party channel either rely on coordinator-like components or make design trade-offs that reintroduce centralized responsibilities (e.g., virtual hubs and coordinator-assisted MPCs). These limitations are evident in representative systems and analyses. \cite{8835315,ye2020garouefficientsecureoffblockchain,10024888,DBLP:journals/iacr/AumayrAM22}

\prot\ provides a fully decentralized, leaderless transaction substrate over multi-party channels: each group operates a jointly funded off-chain ledger (a hyperedge) and members validate updates collectively so that no single node orders, coordinates, or monitors payments. At the protocol level, intra-hyperedge transfers are expressed as jointly signed \emph{DAG nodes} that each proposer orders on its private chain and which are periodically checkpointed by threshold-signed \emph{dag roots}. Inter-hyperedge payments compose value transfer using compact, non-replayable proofs-of-transfer (PoTs) exchanged at connector nodes; conditional transactions bind these artifacts across edges so a source hyperedge releases funds only after the corresponding PoT is verified. §\ref{sec:inter} sketches the composition and correctness argument.

\prot\ targets three practical properties. First, it is \emph{leaderless}: proposer-ordered DAGs allow participants to issue updates in parallel without a single point of control or failure. Second, it is \emph{verifiable and auditable}: threshold-signed DAG roots and PoTs \emph{Proofs-of-Transfer} provide compact cryptographic evidence for off-chain settlements and bounded on-chain exits. Third, it \emph{improves liquidity utilization} by enabling transactions to any member of a hyperedge, reducing the need for explicit rebalancing transactions.

We implemented a prototype and evaluated both intra-hyperedge and inter-hyperedge execution under realistic workloads. In the intra-hyperedge setting, a deployment with 150 participants executing 100{,}000 transactions achieved a transaction success ratio of $94.69\%$ under heavy load i.e high payment size. The observed balance skewness remained low, indicating that there were no signs of liquidity concentration. For the full hypergraph evaluation, we simulated 100 hyperedges with overlapping participants and measured end-to-end inter-hyperedge behavior. Across long runs of up to \(100{,}000\) transactions, the protocol sustained high success rates in the range of \(85\%\) to \(95\%\), while maintaining a low average hop count and exhibiting significantly lower liquidity depletion compared to existing protocols. These results show that \prot\ maintains stable performance under load while avoiding liquidity exhaustion and coordination issues typical of payment-channel networks. Detailed experimental results are presented in Section~\ref{sec:evaluation}, and the corresponding safety and liveness arguments appear in Section~\ref{sec:security}.

In summary, our contributions are as follows:
\begin{itemize}
\item We introduce \prot, a fully decentralized multi-party payment channel network.
\item We provide formal security proofs establishing the correctness and robustness of \prot.
\item We present a concrete implementation of \prot\ along with a comprehensive experimental evaluation.
\end{itemize}

The remainder of the paper is organized as follows. Section~\ref{sec:background} presents the system model. Sections~\ref{sec:intra} and~\ref{sec:inter} describe the intra- and inter-hyperedge payment protocols, respectively. Channel closure and penalties for malicious behavior are introduced in Sections~\ref{sec:closure} and~\ref{sec:penalty}. Section~\ref{sec:security} establishes the liveness and consistency properties of \prot. Section~\ref{sec:evaluation} provides a detailed evaluation of \prot. Future directions and related work are discussed in Sections~\ref{sec:future} and~\ref{sec:related}, respectively. Finally, Section~\ref{sec:conclusion} concludes the paper.


%% file: paper/sections/background_and_model.tex
\section{System Model and Flow}
\label{sec:background}
In this section, we present the system model, describing the structure and operation of the multi-party channel. We then detail the system flow, including how channel state is maintained and updated, and how \prot\ supports parallel transaction execution while ensuring consistent state evolution. Finally, we introduce the notation used throughout the remainder of the paper.


\noindent \textit{\bf Hyperedge funding \& balances:}
In \prot\ set of parties \(P=\{u_1,\dots,u_n\}\) forms a channel by depositing balance \(b_i\) through $in_i$(UTXO). This channel is created by an on-chain funding transaction that consists of $n$ inputs and an output, as depicted below: 
\[
\mathrm{FundingTx}:(in_1,\ldots,in_n)\longrightarrow(out_H),
\]

The output is locked under a multi-party spending policy of \(m\)-of-\(n\) multisignature \footnote{note that one can use different policies such as those coded in a taproot script}, where $m \leq n$. Once a channel is created, several off-chain transfers can take place, post which a channel is closed using a closing transaction that consumes \(out_H\) to pay each participant their final balance \(b_i^T\). Similar to the conventional payment channel network, a party, $p_i$, can be part of several multi-party channels.

Unlike in the conventional payment channel network where only two parties are involved in a channel, which reduces to a graph, \prot\ models set of multiparty channel as a hypergraph and each multi-party channel is modeled as a hyperedge. Formally, Each hyperedge is represented as  \(e=(P,B)\), where \(P=\{u_1,\dots,u_n\}\) is the participant set and \(B=[b_1,\dots,b_n]\) is the balance vector, where \(b_i\) represents \(u_i\)'s contribution in the channel.




\noindent \textit{\bf State commitment:}
In \prot, all participants in a channel must agree on a single balance vector
\(B=[b_1,\dots,b_n]\) which we refer as a {\em channel state}. Each off-chain transaction update the channel state to new state say \(B^T=[b^T_1,\dots,b^T_n]\). Furthermore, channel state and hence a channel at particular time instance is uniquely represented by Merkle root, say \(\widehat{R}\), where leaf of the Merkle tree corresponding to participant \(u_i\) is
\[
L_i = H(\mathit{id}_{u_i} \parallel b_i),
\] 
This Merkle root \(\widehat{R}\) is used as a reference to updated channel state during an off-chain updates. The Merkle root representing the latest updated state is committed on-chain for the final settlement while closing the channel.





\noindent \textit{ \bf DAG node (format and role):} In conventional PCNs, each channel state update—corresponding to the updated balances of the two channel participants—is validated by requiring both parties to jointly sign it. In contrast, in \prot, multiple parties may transact independently and concurrently within the same channel. For example, user \(u_1\) may transact with \(u_2\) while, at the same time, user \(u_4\) transacts with \(u_3\).

A straightforward approach to ensuring consistency in such a setting would be to require every transaction to be endorsed by an honest majority of channel participants. However, this incurs substantial communication overhead, specifically \(O(n)\), where \(n^2\) denotes the number of participants in the channel. Furthermore, this hinders simultaneous transfers initiated by multiple users.

\begin{figure}[h]
    \centering
    \includegraphics[width=0.45\textwidth]{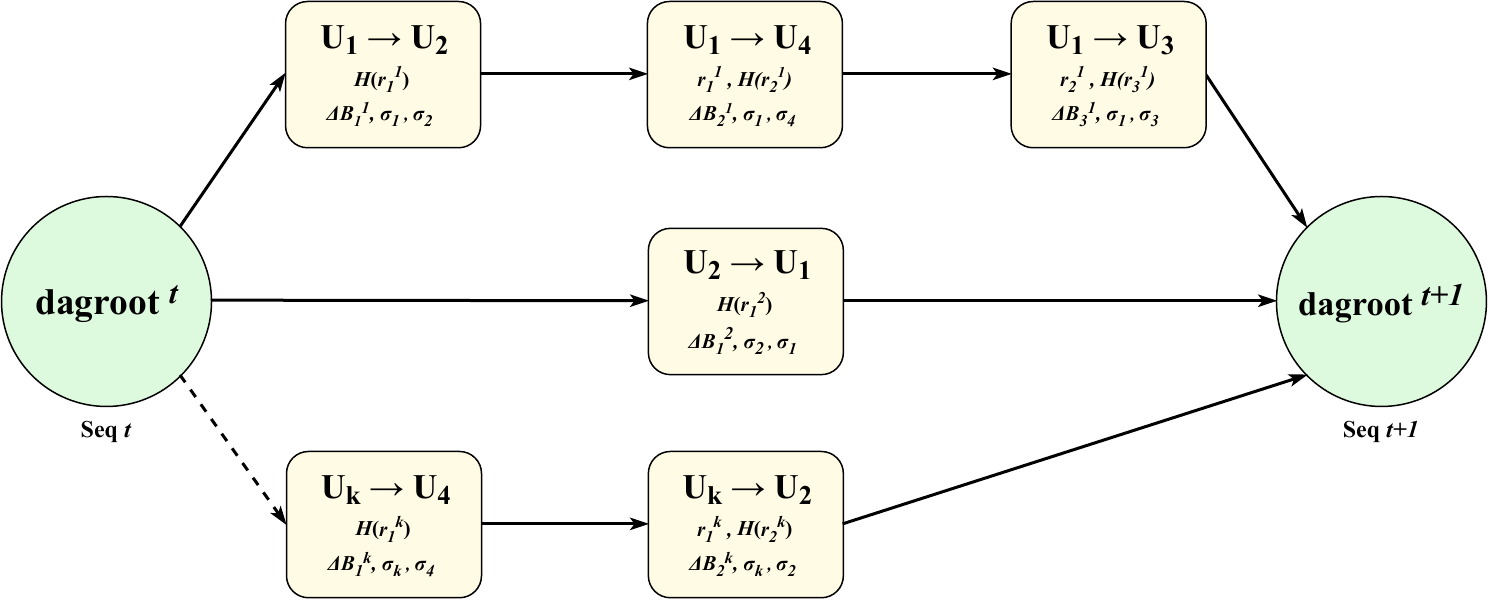}
    \caption{DAG structure inside a hyperedge. Each participant maintains a private proposer chain of DAG nodes; dag roots checkpoint a globally consistent state after threshold endorsement.}
    \label{fig:hyperedge_tx}
\end{figure}

To overcome this, we represent channel state evolution as a novel Directed Acyclic Graph (DAG) structure that each user maintains locally. DAG composed of a sequence of DAG roots and a set of DAG nodes. Each DAG node represents a bilateral transaction between two users, as illustrated in Figure~\ref{fig:hyperedge_tx}. For instance, a DAG node labeled \(u_1 \rightarrow u_2\) denotes a transfer from \(u_1\) to \(u_2\), along with the corresponding local state update.

State updates induced by DAG nodes are not immediately committed. Instead, a state update becomes committed only when the corresponding node is finalized by being appended under a subsequent DAG root. For example, in Figure~\ref{fig:hyperedge_tx}, the state update associated with node \(u_1 \rightarrow u_2\) is committed only after it is appended by \(\mathit{dagroot}^{t+1}\).


With this intuition, we now proceed to the formal definition of DAG nodes. Formally each DAG node consists of following fields:
\[
\mathrm{node}=\langle u_s,u_r,v,f,H(r_{\mathrm{prev}}),H(r_{\mathrm{rev}}),r_{\mathrm{prev}},\sigma_s,\sigma_r\rangle.
\]

Definitions:
\begin{itemize}
  \item \(u_s,u_r\): sender and receiver identifiers
  \item \(v\): proposed transfer amount (the transaction value)
  \item \(f\): fee
  \item \(H(r_{\mathrm{prev}})\): hash of the sender's previous revocation secret (references the latest unrevealed tip)
  \item \(H(r_{\mathrm{rev}})\): hash of the fresh revocation secret for this (current) DAG node
  \item \(r_{\mathrm{prev}}\): the revealed revocation secret that invalidates the referenced previous tip
  \item \(\sigma_s,\sigma_r\): digital signatures of sender and receiver on the DAG node
\end{itemize}

Furthermore, each node is annotated with a balance delta \(\Delta B\), which captures the effect of the transaction on the hyperedge (channel) balance vector. That is, \(\Delta B\) encodes the channel state transition and is applied only when a DAG root finalizes and appends to the corresponding set of nodes. The state update mechanism based on \(\Delta B\) is described in detail in Section~\ref{sec:intra}.

The revocation tuple \((H(r_{\mathrm{prev}}),r_{\mathrm{prev}},H(r_{\mathrm{rev}}))\) enforces a forward-only per-sender chain: revealing \(r_{\mathrm{prev}}\) proves progression past the previous tip, prevents replay of older state, and supports deterministic dispute checks during root finalization or on-chain verification. In other words, revealing \(r_{\mathrm{prev}}\) irrevocably invalidates the preceding state. Whenever a user initiates a new transaction, it discloses the revocation secret associated with its most recent node that has not yet been finalized by a DAG root. For example, in Figure~\ref{fig:hyperedge_tx}, if user \(u_1\) initiates a transaction with \(u_3\) while the latest unfinalized node is \(u_1 \rightarrow u_4\), then \(u_1\) reveals the revocation secret \(r_2^1\). This disclosure invalidates the state represented by \(u_1 \rightarrow u_4\) and permits the creation of a new node \(u_1 \rightarrow u_3\) that extends the DAG.

\textit{\bf Local DAG:}  
In \prot, as stated earlier, each channel participant maintains a local DAG. After executing a transaction, a user appends a new node to its local DAG, invalidates the previous unfinalized state, and broadcasts the node to all channel participants. Upon receipt, each participant validates the node by verifying the signatures of both transacting users, checking the correctness of \(H(r_{\mathrm{prev}})\) and the revealed \(r_{\mathrm{prev}}\), and ensuring that the transaction value \(v\) does not exceed the sender’s remaining channel balance, before appending it to their local DAG copy.


The DAG is organized into separate \emph{proposer chains}, one for each participant. For a participant \(u_i\), only \(u_i\) is allowed to propose nodes on its own chain. Other participants may read or reference that chain, but they cannot extend or modify it.

This restriction prevents a participant from creating two different versions of its own transaction history, because any new node must extend the latest unrevealed tip of its own chain. At the same time, different participants can create nodes on their own chains in parallel.


\textit{\bf DAG roots and finalization:}
While DAG nodes record individual transactions, channel balance changes only when a \emph{DAG root} is finalized. A DAG root acts as a global checkpoint that turns many pending DAG nodes into an agreed state.

Once the number of DAG nodes along any branch reaches a threshold \(k\), a DAG root proposer (explained next) creates a new DAG root that finalizes (commits) all state updates induced by the DAG nodes between the previous DAG root and the newly created one. For example, in Figure~\ref{fig:hyperedge_tx}, \(\mathit{dagroot}^{t+1}\) finalizes the state updates corresponding to the transactions \(u_1 \rightarrow u_2\), \(u_1 \rightarrow u_4\), \(u_1 \rightarrow u_3\), \(u_2 \rightarrow u_1\), \(u_k \rightarrow u_4\), and \(u_k \rightarrow u_1\).


Formally, a DAG node \(\mathit{dagroot}^{t}\)  contains: (i) the previous DAG root (\(\mathit{dagroot}^{t-1}\)), 
(ii) the set of included DAG nodes (or a Merkle commitment to them)- this essential to facilitate cross-channel payment (inter-hyperedge payment—explained in section~\ref{sec:inter}), (iii) Updated state of all the users to be committed, i.e., updated balance vector of all the users which is encoded with a respective Merkle root ($\widehat R$) and (iv) Set of signature that endorse the validity of the DAG root. Please note, a DAG root, say \(\mathit{dagroot}^{t}\), is finalized only if it carries a threshold signature \(\tau_{\mathcal S}(R)\) from a signer set \(\mathcal S\) with \(|\mathcal S|>2n/3\). Together, these checks ensure that no participant’s history is altered and that no minority coalition can finalize an invalid state.




\textit{\bf DAG root proposer selection:} We select user \(u_i\) as proposer with the minimum value of
\(H(\mathit{id}_{u_i} \parallel H(\mathit{dagroot}_{\mathrm{prev}}))\).
The selected proposer is responsible for proposing the next DAG root,
denoted \(\mathit{dagroot}_{\mathrm{cur}}\).
We model the hash function \(H(\cdot)\) as a random oracle. Since the DAG
contents—and hence \(\mathit{dagroot}_{\mathrm{prev}}\)—are jointly
determined by transactions exchanged among multiple users and are not
controlled by any single participant, the proposer selection process is
effectively random and unbiased.



\vspace{0.3em}\noindent\textbf{Quick notation:}
\(e=(P,B)\): hyperedge; \(out_H\): funding output; \(\widehat R\): Merkle root of balances; \(\Delta B\): balance delta; \(r_{\mathrm{prev}},r_{\mathrm{rev}}\): revocation secrets; \(\mathcal S\): threshold signer set.

%% file: paper/sections/intra_hyperedge.tex
\section{Intra-Hyperedge Transactions}
\label{sec:intra}

In this section, we describe how \prot\ facilitates transfers between users within the same channel, i.e., when both parties involved in the transaction belong to a single hyperedge. The objective is to allow multiple participants to submit and validate transfers concurrently while preserving one well-defined balance vector that can be enforced on-chain during channel closure.

The complete process of intra-hyperedge transaction is elaborated in Algorithm~\ref{alg:intra}. We explained the algorithm in two parts. First, construction of transaction node to append it to the DAG. Second, validation and verification of the transaction node to append it to the DAG. \\

\begin{algorithm}[t]
\caption{Intra-Hyperedge Payment Protocol}
\label{alg:intra}
\begin{algorithmic}[1]

\REQUIRE Sender $u_i$, receiver $u_j$, amount $v$, fee $f_i$, previous secret $r_{\mathrm{prev}}$
\ENSURE Valid $\mathrm{dagnode}$ accepted into DAG or rejection

\STATE $u_i$ samples fresh secret $r_{\mathrm{rev}}$
\STATE Let, $h_{\mathrm{rev}} \gets H(r_{\mathrm{rev}})$;  $h_{\mathrm{prev}} \gets H(r_{\mathrm{prev}})$

\STATE Construct node
\[
\ell \gets \langle u_i, u_j, v, f_i, H(r_{\mathrm{prev}}), h_{\mathrm{rev}} \rangle
\]

\STATE $u_i$ signs $\ell$ and sends $(\ell, \sigma_i)$ to $u_j$

\STATE \textbf{At receiver $u_j$:}
\STATE Verify signature $\sigma_i$
\STATE Verify proposer-chain consistency using $H(r_{\mathrm{prev}})$

\STATE $b_i \gets$ sender balance in latest finalized $\mathrm{dagroot}$
\STATE $O_i \gets$ sum of all the $\Delta B$ in the $u_i$'s proposer chain
\STATE $b_i^{\mathrm{avail}} \gets b_i + O_i$

\IF{$b_i^{\mathrm{avail}} < v$}
    \STATE Reject proposal
    \RETURN
\ENDIF

\STATE $u_j$ signs $\ell$ and returns it to $u_i$

\STATE \textbf{At sender $u_i$:}
\STATE Reveal $r_{\mathrm{prev}}$ and broadcast $(\ell, \sigma_i, \sigma_j, r_{\mathrm{prev}})$

\STATE Insert $\ell$ into local DAG

\STATE \textbf{At every participant:}
\STATE Verify $u_i'$s and $u_j'$ signatures and $h_{\mathrm{prev}}$(included in the previous latest transaction node) equals $H(r_{\mathrm{prev}})$
\STATE Insert $\ell$ into local DAG if valid


\end{algorithmic}
\end{algorithm}

\noindent \textit{DAG node construction by the sender:}  
Let \(r_{\mathrm{prev}}\) denote the revocation secret corresponding to \(u_i\)’s most recent uncommitted DAG node, and let \(h_{\mathrm{prev}}\) denote the hash of \(r_{\mathrm{prev}}\) stored in that node. To propose a new transaction, \(u_i\) samples a fresh secret \(r_{\mathrm{rev}}\) and computes \(h_{\mathrm{rev}}=H(r_{\mathrm{rev}})\). The proposed transaction implies a symbolic balance delta
\[
\Delta B = [-v - f_i,\; +v,\; \delta,\ldots,\delta],
\]
where the first two entries debit \(u_i\) and credit \(u_j\). The remaining entries distribute the sender fee among the other \(n-2\) participants. For \(n>2\), \(\delta=f_i/(n-2)\); for \(n=2\), \(\delta=0\).

User \(u_i\) constructs a DAG node
$\ell \gets \langle u_i, u_j, v, f_i, H(r_{\mathrm{prev}}), h_{\mathrm{rev}} \rangle$,
signs it, and sends it to the receiver \(u_j\) (lines~3--4 of Algorithm~\ref{alg:intra}).\\

\noindent \textit{Verification and co-signing at the receiver: }
Upon receiving \(\ell\), user \(u_j\) validates the node by first computing the current balance state up to the latest uncommitted DAG node in its proposer chain. Specifically, \(u_j\) aggregates the balance updates \(\Delta B\) from all uncommitted DAG nodes and applies them to the balance corresponding to the most recent committed DAG root. This ensures that \(u_i\) has sufficient remaining balance to cover the transfer value \(v\) (lines~9--11 of Algorithm~\ref{alg:intra}).

Next, \(u_j\) verifies that \(u_i\) intends to invalidate its previous uncommitted DAG node by checking whether the hash \(H(r_{\mathrm{prev}})\) included in \(\ell\) matches the hash stored in the latest uncommitted DAG node. At this stage, \(u_i\) does not reveal \(r_{\mathrm{prev}}\), since knowledge of this secret would allow any party to invalidate the previous node. Note that the tuple
$\bigl(H(r_{\mathrm{prev}}),\; r_{\mathrm{prev}},\; H(r_{\mathrm{rev}})\bigr)$ embedded in the DAG node enforces forward progress: revealing \(r_{\mathrm{prev}}\) irrevocably invalidates the preceding uncommitted DAG node. Upon successful validation, \(u_j\) signs \(\ell\) and returns it to \(u_i\) (line~15 of Algorithm~\ref{alg:intra}).

After receiving the signed node, \(u_i\) appends the revocation secret \(r_{\mathrm{prev}}\), reconstruct $\ell  \gets \langle u_i, u_j, v, f_i, H(r_{\mathrm{prev}}), H(r_{\mathrm{rev}}), r_{\mathrm{prev}} \rangle$ by adding $r_{\mathrm{prev}}$, and broadcasts \(\ell\) to all other channel participants (line~17 of Algorithm~\ref{alg:intra}).\\

Upon receiving \(\ell\), each other channel participant validates the node by verifying the signatures of both the sender \(u_i\) and the receiver \(u_j\). In addition, they verify the correctness of \(H(r_{\mathrm{prev}})\) and independently validate the balance state using the same procedure as the receiver.

Subsequently, the newly added DAG node \(\ell\) may either be extended by another DAG node representing a subsequent transfer initiated by \(u_i\), or be followed by a DAG root that commits the state update induced by \(\ell\).

As discussed earlier, A root becomes final only when a threshold signer set \(\mathcal S\) with \(|\mathcal S|>2n/3\) produces a threshold signature \(\tau_{\mathcal S}(R)\). The final balance vector added to the DAG root is calculated as $B' = B + \sum_{\ell\in\mathcal L}\Delta B_\ell$, where \(\mathcal L\) is the set of nodes across all the proposer chains between the last DAG root up to the current one.

Having discussed intra-hyperedge transactions, we now turn to inter-hyperedge transaction, which are described next.

%% file: paper/sections/inter_hyperedge.tex
\section{Inter-Hyperedge Transactions}
\label{sec:inter}

In this section we explain how \prot\ moves value between participants that belong to different independently funded hyperedges. 

Consider a sender \(u_b\) in hyperedge \(H_b\) who wishes to transfer value \(v\) to a receiver \(u_a\) in hyperedge \(H_a\). Each hyperedge maintains its own DAG and finalizes state using threshold-signed DAG roots, so the transfer must preserve atomicity across multiple DAGs
while respecting the fact that each hyperedge finalizes state independently via its own DAG root. The transfer proceeds through a common 
participant\footnote{We use the terms ``participant'' and ``user'' interchangeably throughout the paper.} \(u_{ab}\) that is present in both hyperedges; this common participant facilitates inter-hyperedge value exchange.

\prot\ facilitates inter-hyperedge transfers value by introducing \emph{conditional DAG nodes} with \emph{proofs-of-transfer} (PoTs) (discussed in upcoming sub-sections)\footnote{Note that the HTLC used in the conventional PCN wont work here}.

In a nutshell, the inter-hyperedge transfer takes place as follows: The sender issues a conditional DAG node in its own hyperedge and send it to the common participant in the same hyperedge, this promises funds to the common participant, if the common participant can provide a PoT in time \(\mathcal{T}\) (validity of the conditional DAG node) proving that the intended receiver has been paid in the receiver's hyperedge. The common participant first executes the corresponding transaction inside the receiver's hyperedge; once the transaction is committed on the receiver's hyperedge, i.e., appended by the DAG root
the common participant forms a proof-of-transfer (\(\Pi\)). \(\Pi\) consists of the finalized DAG node, the two consecutive threshold-signed DAG roots between which DAG node representing the transaction resides, and the corresponding signature sets.

The common participant then combines \(\Pi\) with the conditional DAG node and broadcasts it to the sender's hyperedge; on successful verification the sender's conditional DAG node is accepted just like a regular DAG node\footnote{Note that a regular DAG node is broad-casted by the sender but in conditional DAG node is broad-casted by the receiver}. 
This execution structure ensures that transfer is done in the sender’s hyperedge only after the corresponding transfer has been finalized in the receiver’s hyperedge, thereby achieving atomicity without requiring global coordination. Each of these steps are illustrated in Algorithm~\ref{alg:inter}.

\begin{figure}[t]
    \centering
    \includegraphics[width=0.3\textwidth]{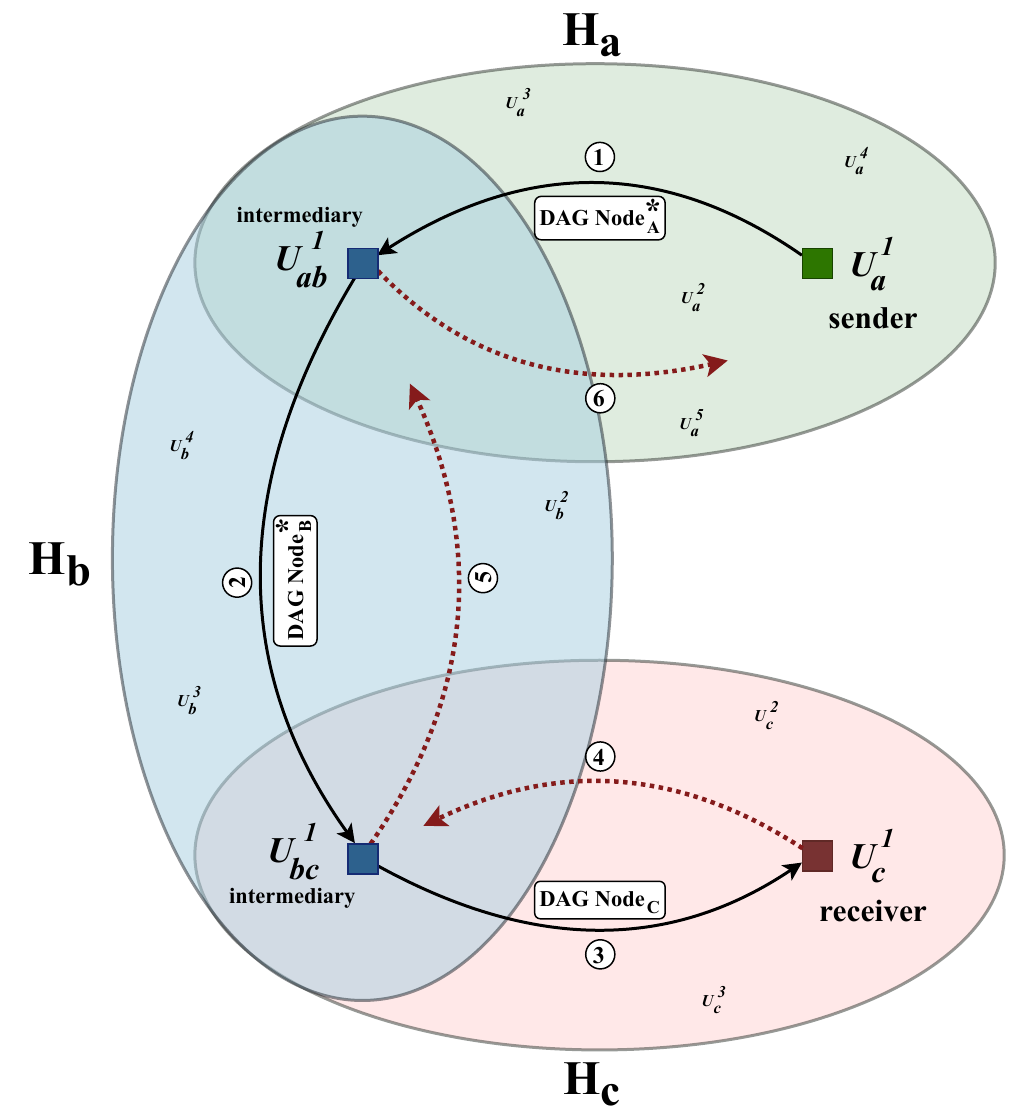}
    \caption{Inter-hyperedge transaction across three independently evolving hyperedges \(H_a\), \(H_b\), and \(H_c\). A transaction initiated by sender \(u^{1}_{a}\) in \(H_a\) is relayed through the intermediary participants \(u^{1}_{ab}\) and \(u^{1}_{bc}\) and finally delivered to the receiver \(u^{1}_{c}\) in \(H_c\). Solid arrows denote forward creation of DAG nodes, while dashed arrows indicate backward propagation of proofs-of-transfer.}
    \label{fig:inter_hyperedge}
\end{figure}

For clarity, we walk through Algorithm~\ref{alg:inter} using the example shown in Figure~\ref{fig:inter_hyperedge}. The example illustrates a transfer from the sender \(u^{1}_{a} \in H_a\) to the receiver \(u^{1}_{c} \in H_c\), where \(H_a\) and \(H_c\) denote the corresponding hyperedges. Since \(H_a\) and \(H_c\) are not directly connected, the transfer must be routed through an intermediate hyperedge \(H_b\). Consequently, the transfer follows the path $u^{1}_{a} \rightarrow u^{1}_{ab} \rightarrow u^{1}_{bc} \rightarrow u^{1}_{c}$. Multiple paths may exist between a sender and receiver across different hyperedges. \prot\ uses BFS to select a shortest path, though it remains agnostic to the specific routing and path-finding algorithms used in the hypergraph~\cite{hpn}.
As discussed earlier, this process is enabled using conditional DAG nodes and Proof of Transfer (PoT). We first describe these two components.



\begin{algorithm}[t]
\caption{Inter-Hyperedge Payment Protocol (Multi-Hop)}
\label{alg:inter}
\begin{algorithmic}[1]

\REQUIRE Path of hyperedges 
$H_a \rightarrow H_b \rightarrow \cdots \rightarrow H_z$  
with connector nodes 
$u_{ab}, u_{bc}, \dots, u_{yz}$;  
sender $u_a \in H_a$, receiver $u_z \in H_z$, amount $v$

\vspace{0.4em}
\STATE \textbf{Phase 1: Forward conditional commitments}
\FORALL{adjacent pairs $(H_x, H_{x+1})$ along the path}
    \STATE Sender-side participant $u_x$ constructs conditional DAG node $\mathrm{Node}^\ast_x$ in $H_x$
    \STATE Node encodes payment $(u_x \rightarrow u_{x,x+1}, v)$
    \STATE Node embeds predicate $\mathcal{P}_{x+1}$:
    \hspace{1em}``a valid proof-of-transfer $\Pi_{x+1}$ from $H_{x+1}$ is provided''
    \STATE Node includes the revocation secret of the sender’s last proposed DAG node
    \STATE $\mathrm{Node}^\ast_x$ is signed and handed to the common participant $u_{x,x+1}$
\ENDFOR

\vspace{0.4em}
\STATE \textbf{Phase 2: Execution at receiver's hyperedge}
\STATE Final common participant $u_{y,z}$ executes ordinary intra-hyperedge payment
$(u_{y,z} \rightarrow u_z, v)$ in $H_z$
\STATE Wait until the node is included between two threshold-finalized dagroots
$\mathrm{dagroot}^{t-1}_z \rightarrow \mathrm{dagroot}^{t}_z$
\STATE Construct proof-of-transfer $\Pi_z$

\vspace{0.4em}
\STATE \textbf{Phase 3: Backward proof propagation}
\FOR{$x = y$ down to $a$}
    \STATE Common participant $u_{x,x+1}$ submits $\Pi_{x+1}$ inside $H_x$
    \STATE Participants verify predicate $\mathcal{P}_{x+1}(\Pi_{x+1})$
    \STATE On success, conditional node $\mathrm{Node}^\ast_x$ is upgraded to a valid DAG node
    \STATE After finalization, construct proof $\Pi_x$ from the new dagroot
\ENDFOR

\vspace{0.4em}
\STATE \textbf{Phase 4: Completion}
\STATE Finalization of $\mathrm{Node}^\ast_a$ completes the transfer
\STATE End-to-end payment $(u_a \rightarrow u_z, v)$ is now irrevocable

\vspace{0.4em}
\ENSURE Atomic settlement across all hyperedges on the path

\end{algorithmic}
\end{algorithm}

\begin{figure*}
    \centering
    \includegraphics[width=0.9\textwidth]{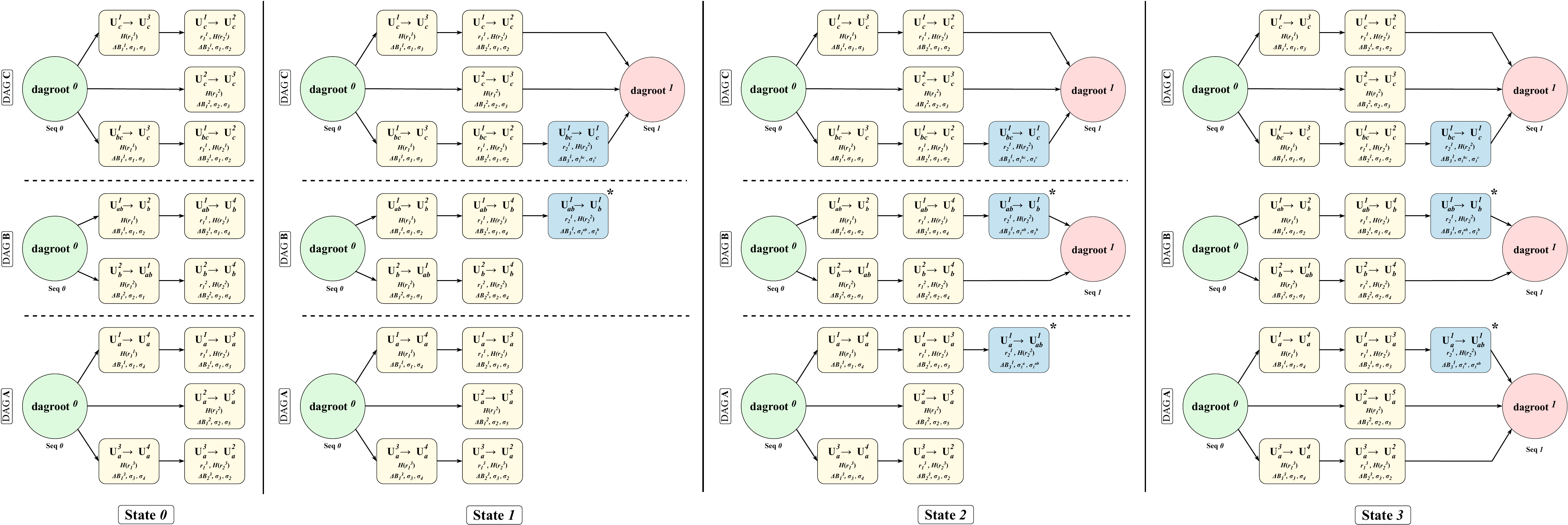}
    \caption{DAG evolution during an inter-hyperedge transaction (States 0--3). Columns are global snapshots (State 0..3); rows show the per-hyperedge DAGs (top: DAG\_C, middle: DAG\_B, bottom: DAG\_A). Blue boxes mark the DAG nodes on the transaction path; yellow boxes show unrelated concurrent activity. Green ovals denote initial (starting) DAG roots and pink ovals denote newly finalized DAG roots produced as the example transaction completes. Numbered arrows indicate forward placement of conditional DAG nodes (1--3) and backward propagation of proofs-of-transfer (4--6).}
    \label{fig:dags_example}

\end{figure*}

\subsection*{Conditional Commitment in \(H_a\)}
\label{sec:inter:cond}

The protocol begins in the sender’s hyperedge \(H_a\). Consider the example in
Figure~\ref{fig:inter_hyperedge} where sender \(u^{1}_{a}\in H_a\) wishes to
pay receiver \(u^{1}_{c}\in H_c\). The first concrete action is a conditional commitment issued by the sender (see Algorithm~\ref{alg:inter}, Phase 1, lines 6–9). \(u^{1}_{a}\) constructs a conditional DAG
\(\mathrm{Node}^\ast_A\) that promises \((v+\delta)\) to the common participant
\(u^{1}_{ab}\)\footnote{Note that there can be multiple common users. Sender randomly select one to facilitate the transfer} and binds this promise to a predicate \(\mathcal{P}_b(v)\) (Algorithm~\ref{alg:inter}, line 5) (asserting that a corresponding transaction to \(u^{1}_{bc}\) will be finalized
in \(H_b\).

Similarly, \(u^{1}_{ab}\) creates and shares a conditional DAG node \(\mathrm{Node}^{\ast}_{B}\) with a participant $u^{1}_{bc}$, that is common to hyperedges \(H_b\) and \(H_c\) (Algorithm~\ref{alg:inter}, lines 3–7). This node promises a value of \((v+\delta)\) to the ultimate receiver \(u^{1}_{c}\). Moreover, it cryptographically binds this promise to \(\mathcal{P}_c(v)\), asserting that the corresponding transaction to \(u^{1}_{c}\) will be finalized in \(H_c\).

The conditional DAG node’s structure follows the standard DAG node format and
additionally carries the predicate:
\[
\langle u^{1}_{a},\; u^{1}_{ab},\; v+\delta,\; \mathcal{P}_b(v),\;
H(r_{\mathrm{rev}}),\; r_{\mathrm{prev}},\; H(r_{\mathrm{prev}})\rangle.
\]

As with ordinary DAG nodes, the conditional node is signed by the sender  and includes revocation metadata that links it to the sender’s proposer chain (i.e. the chain of the nodes that are proposed by them). Practically, the signed conditional node is sent to the common participant
\(u^{1}_{ab}\) (Algorithm~\ref{alg:inter}, line 6-7)
. From the sender’s perspective this action is binding: revealing the
next revocation secret would revoke \(\mathrm{Node}^\ast_A\), so the sender is
cryptographically committed to the inter-hyperedge transfer until either the
condition is satisfied or a timeout \(\mathcal{T}\) expires. If the common participant
fails to provide the required proof within \(\mathcal{T}\), the conditional
node becomes invalid and the sender may safely resume further transactions on its proposer chain. 

\subsection*{Execution at the Receiver's Hyperedge and Proof-of-Transfer}
\label{sec:inter:pot}

The conditional node in \(H_a\) does not itself move funds. To fulfill the
predicate \(\mathcal{P}_c(v)\), the last-hop common participant \(u^{1}_{bc}\) must execute a regular intra-hyperedge transaction in \(H_c\) (see ~\ref{alg:inter} line 10) that transfers value \(v\) from \(u^{1}_{bc}\) to \(u^{1}_{c}\). Concretely, \(u^{1}_{bc}\) creates a standard DAG \(\mathrm{Node}_C: u^{1}_{bc}\!\rightarrow\!u^{1}_{c}:v\) and propagates it within \(H_c\) (see ~\ref{alg:inter} line 10). When that node is included between two DAG roots, $\mathrm{dagroot}_{c}^{t-1}$ and $\mathrm{dagroot}_{c}^{t}$, the transfer becomes irreversible (see ~\ref{alg:inter} line 11).

At that point \(u^{1}_{bc}\) constructs a proof-of-transfer \(\Pi_c\):
\[
\begin{aligned}
\Pi_c = (&\mathrm{Node}_C,\; \mathrm{dagroot}_{c}^{t-1},\;
\mathrm{dagroot}_{c}^{t},\;
\{\sigma_k\}_{k\in\mathcal{S}_c^{(t-1)}}, \\
&\{\sigma_k\}_{k\in\mathcal{S}_c^{(t)}}).
\end{aligned}
\]

\(\Pi_c\) binds the signed DAG node to the two consecutive
threshold-signed DAG roots and the corresponding signer sets 
Any participant in another hyperedge can verify \(\Pi_c\) by checking the node signature,
verifying the linkage between the two roots, and validating the threshold
signatures. 


\subsection*{Conditional Release in \(H_b\)}
\label{sec:inter:release}

Once \(u^{1}_{bc}\) obtains \(\Pi_c\), it forwards the proof to the upstream
common participant \(u^{1}_{ab}\) (see ~\ref{alg:inter} line 15). The participants of \(H_b\) independently verify
\(\Pi_c\) (see ~\ref{alg:inter} line 16). Upon successful verification, the conditional DAG node issued by
\(u^{1}_{ab}\) (the DAG \(\mathrm{Node}^\ast_B\) placed earlier in \(H_b\)) is
upgraded to a normal DAG node and becomes eligible for inclusion in the next
threshold-finalized DAG root of \(H_b\) (see ~\ref{alg:inter} line 17) . The upgrade embeds fresh revocation
values and commits the balance effects inside \(H_b\).

After \(H_b\) finalizes the upgraded node, \(u^{1}_{ab}\) constructs
\(\Pi_b\) for the now-finalized node in \(H_b\) and forwards \(\Pi_b\) to
\(H_a\) (see ~\ref{alg:inter} line 18). \(H_a\) verifies \(\Pi_b\) and, on success, upgrades and finalizes
its original conditional DAG \(\mathrm{Node}^\ast_A\). This backward-chain of
proofs ensures that each upstream conditional node is finalized only after the
downstream transfer it depends on has already been finalized.

\subsection*{Example walk-through (mapping to the diagrams)}
\label{sec:inter:walk}

Figures~\ref{fig:inter_hyperedge} and~\ref{fig:dags_example} visualize the
example above. In the figure the forward path is
\(u^{1}_{a}\rightarrow u^{1}_{ab}\rightarrow u^{1}_{bc}\rightarrow u^{1}_{c}\)
and the proof propagation follows the reverse direction. Fig.~\ref{fig:dags_example}
shows four global snapshots (States 0--3) that make the timing explicit:

State 0: each hyperedge contains independent background transactions (yellow
nodes) and the initial DAG roots are shown in green. No conditional nodes for
this transfer have been finalized.

State 1: the sender \(u^{1}_{a}\) has issued DAG \(\mathrm{Node}^\ast_A\) and the
common participant \(u^{1}_{ab}\) has DAG \(\mathrm{Node}^\ast_B\) in \(H_b\).
The final hop \(\mathrm{Node}_C\) (in \(H_c\)) is present but not yet
finalized.

State 2: the final hop \(\mathrm{Node}_C\) is included between two
threshold-finalized DAG roots in \(H_c\) (pink root). \(u^{1}_{bc}\)
constructs \(\Pi_c\) and forwards it to \(H_b\).

State 3: verification of \(\Pi_c\) upgrades and finalizes DAG \(\mathrm{Node}^\ast_B\)
in \(H_b\), which yields \(\Pi_b\). After \(H_a\) verifies \(\Pi_b\), DAG
\(\mathrm{Node}^\ast_A\) is upgraded and finalized. The pink roots in each
row of Fig.~\ref{fig:dags_example} show the newly finalized DAG roots that
commit the transfer step-by-step.

\subsection*{Atomicity, Timeouts, and Liveness}
\label{sec:inter:atomicity}

Atomicity follows from two facts. First, conditional DAG nodes cannot be
finalized without a valid PoT; a common participant cannot claim upstream funds unless
it has produced the downstream PoT. Second, PoTs are bound to specific nodes
and adjacent threshold-signed DAG roots, so a PoT cannot be replayed in a
different context.

To avoid indefinite lockup, conditional nodes include a timeout \(\mathcal{T}\).
If the required PoT does not arrive before \(\mathcal{T}\), the conditional
node is treated as invalid and normal proposer-chain progress resumes. In the
timeout case honest participants recover funds using the on-chain dispute and
closure mechanisms described in Section~\ref{sec:closure}.

%% file: paper/sections/settlement_and_balance_update.tex
\section{Penalty Mechanism}
\label{sec:penalty}

\subsection*{Conditional-Node Equivocation}
\label{sec:penalty:case1}

Upon providing a signed conditional DAG node to a common participant for an inter-hyperedge transfer, the sender reveals the previous revocation secret on its proposer chain. Consequently, the common participant holds a sender-signed conditional DAG node and a valid revocation secret, cryptographically binding the sender to the commitment for a fixed interval \(\mathcal{T}\).

During the interval \(\mathcal{T}\), the sender must not publish another DAG node on the same proposer chain. Any violation allows the common participant to combine the signed conditional DAG node with the revealed revocation secret to produce a proof of equivocation, which can be submitted on-chain to trigger punitive resolution (e.g., distribution or burning of the sender’s locked funds) under the closure rules.

This mechanism ensures that once a conditional commitment is issued, the sender is cryptographically prevented from invalidating it by attempting to overwrite or extend its proposer chain during the validity window.

\subsection*{Revocation Violation via Double-Referencing}
\label{sec:penalty:case2}

In intra-hyperedge execution, each participant’s proposer chain enforces a strict revocation-based ordering: once the revocation secret of a DAG node has been revealed, that node can no longer be safely used as a parent for any future DAG node. This property is necessary to ensure that participants cannot reuse or resurrect outdated states.

If a sender nevertheless signs two distinct DAG nodes that both reference the same parent DAG node whose revocation secret has already been revealed, this constitutes a cryptographically provable violation of the proposer-chain rules. The sender’s signatures on both DAG nodes, together with the already revealed revocation secret of the shared parent, form a self-contained proof that the sender attempted to equivocate.

Such a proof can be submitted on-chain and enables punitive resolution under the closure mechanism, including burning or redistribution of the sender’s locked funds. This ensures that participants are economically disincentivized from attempting to bypass the revocation-based safety guarantees of the DAG structure.

%% file: paper/sections/channel_closure.tex
\section{Channel Closure}
\label{sec:closure}

Channel closure returns a hyperedge to the blockchain by settling the balances
committed in the most recent finalized DAG root. The protocol supports three
closure modes—cooperative closure, unilateral dispute-driven closure, and
atomic hyperedge reconfiguration—each designed to preserve safety while
accommodating different failure scenarios.

\subsection*{Cooperative Closure}

In the cooperative case, participants jointly close the hyperedge by submitting
the latest finalized \(\mathrm{DAG\ root}\) together with its proposer
endorsements and supermajority threshold signature (\(>2n/3\)). Since the root
already commits to all parent tips and the corresponding balance vector, each
party can locally verify correctness before authorizing closure.

Once verified, the participants co-sign a single on-chain transaction that
spends the funding output and distributes funds according to the balances
encoded in the finalized root. This mode requires minimal on-chain interaction
and represents the fastest and most efficient closure path.

\subsection*{Unilateral Closure}

If cooperation is not possible, due to unresponsiveness or adversarial
behavior, any honest participant may unilaterally initiate closure by submitting
their latest finalized \(\mathrm{DAG\ root}\) on-chain.

The contract opens a dispute window during which other participants may present
a strictly newer finalized root. Validity is determined by checking proposer
endorsements, parent-root consistency, and the threshold signature. If no newer
root is submitted within the dispute window, the posted root is accepted as
final and balances are settled accordingly.

This mechanism guarantees liveness: even under partial synchrony or adversarial
conditions, funds can always be recovered without coordination.

\subsection*{Atomic Hyperedge Reconfiguration}

Beyond full closure, the protocol supports \emph{atomic hyperedge
reconfiguration}, allowing a participant to exit while the remaining parties
continue operating without interruption.

Let \(u_{\mathsf{exit}}\) be a participant wishing to leave. Reconfiguration is
performed through two covenant-linked transactions that execute atomically.

Step 1 (Exit Transaction).
The exiting participant submits the latest finalized \(\mathrm{DAG\ root}_t\),
its threshold signatures, and a Merkle proof for its balance
\(b_{\mathsf{exit}}^{(t)}\). This transaction withdraws
\(b_{\mathsf{exit}}^{(t)}\) from the funding output.

Step 2 (Reconfiguration Transaction).
A second transaction, whose hash is committed by the first, creates a new
funding output for the remaining participant set
\(
H' = H \setminus \{u_{\mathsf{exit}}\}.
\)
The new hyperedge is initialized with balance vector $B^{(t)}_{-{\mathsf{exit}}}
= (b_i)_{u_i \in H'}$
and commitment
$\widehat{R}'_0 = \mathrm{MerkleRoot}(B^{(t)}_{-{\mathsf{exit}}})$.

The two transactions are mutually dependent: the exit transaction is valid only
if the reconfiguration transaction executes, and the reconfiguration transaction
is valid only if the exit is confirmed. This guarantees atomicity.

As a result, a participant can safely leave without halting the hyperedge or
requiring coordination from the remaining parties, while the remaining users
continue with a correctly initialized channel state.





%% file: paper/sections/security_analysis.tex
\section{Security Analysis}\label{sec:security}

This section outlines the security properties, in terms of consistency and liveness of \prot\ and the assumptions under which they hold. 

We consider a distributed system consisting of \(n = 3f+1\) \emph{hyperedge participants}, of which at most \(f\) may exhibit Byzantine faults. Participants communicate over authenticated channels using cryptographic mechanisms such as digital signatures, preventing spoofing and message forgery. The network operates under partial synchrony: there exists an unknown Global Stabilization Time (GST) after which message delivery and processing delays are bounded by a known constant \(\Delta\), while prior to GST delays may be arbitrary. All honest hyperedge participants output the same state evolution for identical ordered inputs. Furthermore, we assume that identity of all the participants of \prot\ is publicly known.\\

\noindent{\bf Consistency:} We prove that all honest participants agree on the identity of the DAG root proposer. In particular, we show that no two honest participants can commit to different sets of DAG nodes with different DAG roots proposed by distinct proposers. Consequently, the committed DAG has a unique DAG root proposer agreed upon by all honest participants. 

As discussed earlier, the DAG root proposer is a participant with the minimum value of
\(H(\mathit{id}_{u_i} \parallel H(\mathit{dagroot}_{\mathrm{prev}}))\), our aim is to prove that every honest participant computes the same minimum value. 

\begin{theorem}
    All honest participants compute the same DAG root proposer
\end{theorem}
\begin{proof}
    All participant identities and hyperedge memberships are publicly known; hence all honest participants agree on the candidate set for DAG root proposer election. Since the hash function \(H\) is deterministic and collision-resistant, it induces a unique ordering with overwhelming probability. Thus, it suffices to show that honest participants agree on the same previous DAG root \(\mathit{dagroot}_{\mathrm{prev}}\).

A valid DAG root requires endorsements from a quorum of at least \(2f+1\) participants. For any two quorum sets \(Q_1\) and \(Q_2\),
\[
\lvert Q_1 \cap Q_2 \rvert 
\ge (2f+1) + (2f+1) - (3f+1) 
= f+1 .
\]
Since at most \(f\) participants are Byzantine, the intersection contains at least one honest participant, who cannot endorse two distinct DAG roots. Therefore, conflicting DAG roots cannot be valid, and all honest participants elect the same DAG root proposer.

\end{proof}

\noindent{\bf Liveness:}
As discussed earlier, When the number of DAG nodes on any proposer chain reaches the threshold \(k\), the DAG root proposer is expected to create a new DAG root. This assumes an honest proposer; in \prot, a malicious DAG root proposer may withhold the DAG root, thereby delaying the commitment of all DAG-node transactions since the previous root.

To handle this scenario, if the number of DAG nodes in any proposer chain exceeds \(k+y\), where \(y\) is a user-defined  parameter that controls the desired progression rate of the chain, and no DAG root is generated, \prot\ triggers the election of a new DAG root proposer. The new proposer is selected as the participant with the next minimum value of
$H(\mathit{id}_{u_i} \parallel H(\mathit{dagroot}_{\mathrm{prev}}))$. We refer to a span of \(y\) DAG nodes as a \emph{round}.

Furthermore, a malicious DAG root proposer may deliberately exclude certain DAG nodes. These excluded nodes will subsequently be considered by the next honest DAG root proposer; however, the corresponding user must re-initiate the transaction. So, to ensure liveness, we have to prove that a honest DAG root proposer must eventually get elected with overwhelming probability, which we will prove next in Theorem~\ref{thm:liveness}.

\begin{theorem}
    \label{thm:liveness}
    An honest DAG root proposer is elected within \(O(\log(1/\varepsilon))\) rounds with probability at least \(1 - \varepsilon\).
\end{theorem}
\begin{proof}

The probability that the elected DAG root proposer is dishonest is
$q \;=\; \frac{f}{n} \;<\; 1$, and consecutive DAG root proposer elections are independent.
Let \( E_T \) denote the event that no honest DAG root proposer is elected in the first \( T \) rounds.

By independence of elections across rounds,
$\Pr[E_T] \;=\; q^T$.
Using the identity \( a^T = e^{T \ln a} \) for \( a \in (0,1) \),
\(
\Pr[E_T] \;=\; e^{T \ln q}.
\)
Since \( 0 < q < 1 \), we have \( \ln q < 0 \). Define
\(
c \;:=\; -\ln q \;=\; \ln\!\left(\frac{n}{f}\right) \;>\; 0.
\)
Hence,
\(
\Pr[E_T] \;=\; e^{-cT}.
\)
This shows that the failure probability decreases exponentially in \( T \).

Let \( \varepsilon > 0 \) be a target failure probability. We want
$\Pr[E_T] \;\le\; \varepsilon$.
Substituting,
\(
e^{-cT} \le \varepsilon
\;\;\Longleftrightarrow\;\;
T \ge \frac{1}{c}\ln\!\left(\frac{1}{\varepsilon}\right).
\)
Thus, $T \;=\; O\!\left(\log \frac{1}{\varepsilon}\right)$ rounds suffice to elect an honest DAG root proposer with probability at least \( 1-\varepsilon \).

\end{proof}



%% file: paper/sections/evaluation.tex
\section{Evaluation}
\label{sec:evaluation}

We evaluate the intra-hyperedge transfer in isolation to establish a baseline for correctness and stability under sustained concurrency. We describe the experimental setup, followed by the evaluation parameters and metrics—some specific to the intra-hyperedge setting—and conclude with results and observations.




\subsection{Experimental Setup}

We have evaluated \prot\ with the simulated network of $\approx$15k nodes (users), which are uniformaly distributed to form a hyperedge with 150 users and hence 100 different hyperedges. 
  
We construct a synthetic hypergraph consisting of $100$ hyperedges, each
containing $150$ participants using snowball sampling~\cite{rebal3}.  
Between adjacent hyperedges, we introduce $8$--$20$ overlapping participants.
As we know, these overlapping nodes serve as intermediaries and are therefore the primary
points where liquidity stress and depletion can occur.

We evaluate two workloads:
\begin{itemize}
  \item \textbf{Short-term run:} $50{,}000$ transactions.
  \item \textbf{Long-term run:} $100{,}000$ transactions.
\end{itemize}

This allows us to distinguish between temporary stability and genuine
long-term robustness.
Each transaction amount corresponds to roughly $5$--$10\%$ of the sender’s balance.  
This is substantially larger than the typical payment sizes observed in Lightning, where transfers are usually small relative to channel capacity (see \cite{coinspaid-lightning-stats, bitcoinvisuals-lightning-capacity}).

Transactions are processed continuously, and system state is periodically (after every 1k transactions) finalized via $\mathrm{dagroot}$ construction.  
At each finalized root, participants update their balance vectors from a DAG root checkpoint.  
The simulator records accepted and rejected transactions, finalized balances, and DAG state after each root.

\subsection{Metrics and Parameters}
We evaluate \prot by varying several key parameters, including the runtime index (e.g., iteration count or DAG root index), which captures specific time instances during the experimental run. In addition, we vary the payment size—representing the transaction value—by an order of magnitude.

Furthermore, to evaluate \prot\ for intra-hyperedge transfers we used two metrics-Transaction Success Ratio (TSR) and Balance Skewness. For inter-hyperedge transfer we used four metrics: two comparative (against Lightning, Shaduf, and Revive)
and two specific to hypergraph execution. Each of these are explained below:

\paragraph{Success Ratio (comparative).}
This metric measures the proportion of successful transactions among all submitted transactions. We evaluate the end-to-end transaction success ratio over time and compare it with reported results for Lightning, Shaduf, and Revive, capturing network reliability under sustained load.

\paragraph{Balance   Skewness} To quantify liquidity dispersion within a hyperedge, we measure \emph{balance skewness} after each finalized $\mathrm{dagroot}$. Let $B^{(t)} = (b^{(t)}_1,\dots,b^{(t)}_n)$ denote participant balances after root $t$, with mean $\bar{b}^{(t)}$. Skewness is defined as $S\bigl(B^{(t)}\bigr)
=
\frac{1}{n}
\sum_{i=1}^n
\frac{\lvert b^{(t)}_i - \bar{b}^{(t)} \rvert}{\bar{b}^{(t)}}$, which measures deviation from uniform liquidity. The maximum value,
$S_{\max}(n) = \frac{2(n-1)}{n}$, occurs when a single participant holds all liquidity.

For $n=150$, $S_{\max}(150) \approx 1.986$.  
Across our experiment, the maximum observed skewness was
$S_{\text{obs}}^{\max} \approx 0.70$, indicating moderate dispersion but far from pathological concentration.

\paragraph{Depletion Count (comparative)}
We measure the number of intermediaries that fail to forward payments due to balance exhaustion. This mirrors channel depletion in PCNs and provides a direct measure of structural fragility.

\paragraph{Liquidity Retention Factor (LRF).}
To quantify the sustainability of overlapping participants, we define the
\emph{Liquidity Retention Factor} (LRF) for each intermediary $u$ as $\mathrm{LRF}_u = \frac{b_u^{\text{current}}}{b_u^{\text{initial}}}$.
LRF measures how much usable liquidity an intermediary retains relative to its
starting balance.  
Values near $1$ indicate sustained usability, while values approaching $0$
indicate effective depletion.  
This is a hypergraph-specific metric and therefore reported only for \prot.


\paragraph{Average Hop Length}
For each successful inter-hyperedge payment, we record the number of hyperedges
traversed along the route, denoted by $\ell$.  
We group transactions by transfer amount $v$ and compute the average hop length as
\(
\bar{\ell}(v) =
\frac{1}{|\mathcal{T}_v|}
\sum_{i \in \mathcal{T}_v} \ell_i,
\)
where $\mathcal{T}_v$ is the set of all successful transactions with amount $v$
and $\ell_i$ is the hop length of transaction $i$.

This metric captures whether larger payments require longer routes
(more intermediaries) or whether routing depth remains stable across value scales.
Unlike success and depletion, this reflects structural efficiency rather than
failure resilience.

\subsection{Results and Observations}
We will first elaborate on  our observation for intra-hyperdge evaluation which is followed by discussion on inter-hyperedge. \\

\noindent{\bf Intra-hyperedge evaluation: }
As depicted in Figure~\ref{fig:combined} the TSR varies between 91\% to 96\% across the run of our experiment. Specfically, for the $100{,}000$ attempted transactions, $94{,}690$ transactions
succeeded while $5{,}310$ failed, yielding an aggregate success ratio of
$94.69\%$. The minimum observed success ratio over all finalized roots was
$90.8\%$, while several roots achieved $100\%$ acceptance.

All observed failures arise exclusively from insufficient sender balance at
proposal time. No failures originate from routing, locking, timeout, or
coordination artifacts, since such mechanisms do not exist within a hyperedge.




Furthermore, as shown in Figure~\ref{fig:combined}\footnote{To the best of our knowledge, there are no large-scale practical experimental studies of MPC-style multi-party payment systems under sustained load, making direct empirical baselines unavailable.}, the balance skewness converges to a saturation level of approximately \(0.6\) after around 20 DAG roots. This initial increase in skewness arises from the redistribution of balances toward a subset of users. Beyond this point, even as additional transactions are processed, the skewness remains largely stable with only minor fluctuations. This behavior indicates that the intra-hyperedge protocol reliably sustains transaction throughput under heavy load without introducing instability, coordination failures, or pathological liquidity collapse.\\


\begin{figure}[t]
\centering
\begin{tikzpicture}

\begin{axis}[
    width=0.38\textwidth,
    height=3.6cm,
    xlabel={DAG Root Index},
    ylabel={Success Ratio (\%)},
    ylabel style={
      at={(axis description cs:-0.15,0.88)},
      anchor=east,
      font=\scriptsize
    },
    xmin=1, xmax=100,
    ymin=85, ymax=101,
    grid=both,
    grid style={gray!25},
    tick label style={font=\scriptsize},
    label style={font=\scriptsize},
    name=leftaxis,
    clip=false
]

\addplot[
    blue,
    mark=none,
    mark size=0.6pt
] coordinates {
(1,100.0)(2,100.0)(3,99.7)(4,99.0)(5,98.4)(6,96.9)(7,96.4)(8,96.1)
(9,95.3)(10,96.1)(11,98.2)(12,96.8)(13,93.9)(14,94.1)(15,95.5)
(16,95.8)(17,95.3)(18,92.2)(19,94.4)(20,96.5)(21,95.1)(22,96.0)
(23,96.0)(24,94.3)(25,93.3)(26,94.7)(27,94.3)(28,93.9)(29,90.8)
(30,92.9)(31,93.0)(32,93.7)(33,96.5)(34,95.5)(35,93.3)(36,95.0)
(37,95.2)(38,94.3)(39,92.2)(40,93.0)(41,93.1)(42,93.1)(43,93.0)
(44,94.5)(45,94.2)(46,92.2)(47,92.9)(48,94.9)(49,94.2)(50,92.7)
(51,95.7)(52,93.8)(53,94.3)(54,93.0)(55,94.3)(56,94.5)(57,91.5)
(58,94.4)(59,96.4)(60,93.5)(61,94.1)(62,94.0)(63,96.1)(64,96.2)
(65,96.3)(66,95.0)(67,93.0)(68,95.9)(69,90.9)(70,93.7)(71,94.0)
(72,91.6)(73,92.3)(74,94.9)(75,95.9)(76,95.3)(77,94.8)(78,93.5)
(79,92.8)(80,92.9)(81,92.5)(82,91.5)(83,92.9)(84,93.0)(85,95.8)
(86,94.7)(87,95.3)(88,94.7)(89,93.9)(90,94.2)(91,94.2)(92,94.5)
(93,95.7)(94,96.1)(95,96.7)(96,98.1)(97,96.2)(98,96.3)(99,95.5)
(100,94.2)
};

\end{axis}

\begin{axis}[
    width=0.38\textwidth,
    height=3.6cm,
    xmin=1, xmax=100,
    ymin=0.15, ymax=0.80,
    axis y line*=right,
    axis x line=none,
    ylabel={Balance Skewness},
    ylabel style={
      at={(axis description cs:1.15,0.88)},  
      anchor=west,
      font=\scriptsize,
      rotate=180
    },
    tick label style={font=\scriptsize},
    label style={font=\scriptsize},
    name=rightaxis,
    clip=false
]

\addplot[
    red,
    mark=none,
    mark size=0.6pt
] coordinates {
(1,0.182094)(2,0.259672)(3,0.320660)(4,0.365691)(5,0.403693)
(6,0.444829)(7,0.472981)(8,0.517210)(9,0.527666)(10,0.539500)
(11,0.563291)(12,0.606915)(13,0.616760)(14,0.608245)(15,0.596771)
(16,0.599556)(17,0.618014)(18,0.629643)(19,0.614283)(20,0.633287)
(21,0.644549)(22,0.640572)(23,0.641920)(24,0.624096)(25,0.657727)
(26,0.646474)(27,0.667250)(28,0.673814)(29,0.676112)(30,0.672587)
(31,0.680521)(32,0.688338)(33,0.682116)(34,0.686874)(35,0.690424)
(36,0.689127)(37,0.692215)(38,0.702512)(39,0.705609)(40,0.708937)
(41,0.705104)(42,0.699402)(43,0.694331)(44,0.690003)(45,0.693899)
(46,0.695553)(47,0.689980)(48,0.684111)(49,0.684033)(50,0.696576)
(51,0.685608)(52,0.689447)(53,0.702243)(54,0.706180)(55,0.714029)
(56,0.718111)(57,0.722370)(58,0.719771)(59,0.712887)(60,0.700885)
(61,0.693436)(62,0.698089)(63,0.705761)(64,0.708915)(65,0.713301)
(66,0.719654)(67,0.723960)(68,0.728431)(69,0.731768)(70,0.732554)
(71,0.726988)(72,0.729036)(73,0.724237)(74,0.721188)(75,0.718990)
(76,0.720734)(77,0.725540)(78,0.733872)(79,0.713053)(80,0.680855)
(81,0.711811)(82,0.714442)(83,0.718963)(84,0.721986)(85,0.699239)
(86,0.706511)(87,0.711582)(88,0.707479)(89,0.701284)(90,0.698676)
(91,0.697410)(92,0.692501)(93,0.690262)(94,0.689017)(95,0.684971)
(96,0.679033)(97,0.599680)(98,0.620593)(99,0.623792)(100,0.656785)
};

\end{axis}

\node[anchor=south] at ($(leftaxis.north)!0.5!(rightaxis.north)$) {%
  \tiny
  \begin{tabular}{@{}l@{\ \ }l@{}}
    \tikz\draw[fill=blue]  (0,0) rectangle (0.8ex,0.9ex); & Success Ratio \\
    \tikz\draw[fill=red]   (0,0) rectangle (0.8ex,0.9ex); & Balance Skewness
  \end{tabular}%
};

\end{tikzpicture}

\caption{Success ratio and balance skewness across 100 DAG roots.}
\label{fig:combined}
\end{figure}


\noindent{\bf Inter-hyperedge evaluation: } We now evaluate the complete protocol over a full hypergraph topology, where multiple hyperedges coexist and overlapping participants act as intermediaries for inter-hyperedge payments.  
This experiment captures the emergent behavior of the full system, including success under routing pressure, intermediary sustainability, and long-term liquidity dynamics.

\newcounter{subfig}

\begin{figure}
\centering

\makebox[\columnwidth][c]{%
\tiny
\begin{tabular}{@{}l@{\hspace{0.6em}}l@{\hspace{1.0em}}l@{\hspace{0.6em}}l@{\hspace{1.0em}}l@{}}

\tikz\draw[blue, densely dashed, line width=0.7pt] (0,0)--(1.1em,0); COALESCE &
\tikz\draw[red, line width=0.7pt] (0,0)--(1.1em,0); Lightning &
\tikz\draw[green!60!black, line width=0.7pt] (0,0)--(1.1em,0); Shaduf &
\tikz\draw[purple, line width=0.7pt] (0,0)--(1.1em,0); Revive &
\tikz\draw[orange, line width=0.7pt] (0,0)--(1.1em,0); Horcrux \\

\tikz\draw[violet, line width=0.7pt] (0,0)--(1.1em,0); Ticket To Ride &
\tikz\draw[yellow!80!black, line width=0.7pt] (0,0)--(1.1em,0); Speedy Murmurs &

\end{tabular}
}

\resizebox{0.98\columnwidth}{!}{

\begin{tabular}{cc}

\refstepcounter{subfig}\label{fig:success_vs_time}
\begin{minipage}{0.48\columnwidth}
\centering
\begin{tikzpicture}
\begin{axis}[
    width=4.6cm,
    height=3.6cm,
    xlabel={Iteration},
    ylabel={Success (\%)},
    ymin=0.40,
    ymax=1.05,
    ytick={0.40,0.50,0.60,0.70,0.80,0.90,1.00},
    yticklabels={40,50,60,70,80,90,100},
    grid=both,
    grid style={gray!20},
    tick label style={font=\scriptsize},
    label style={font=\scriptsize},
    legend style={font=\scriptsize, at={(0.5,1.12)}, anchor=south, draw=none},
    clip=false
]
\addplot[semithick, blue, densely dashed]
table[col sep=comma,x index=0,y index=5]{final_success_ratio.csv};

\addplot[thin, red]
table[col sep=comma,x index=0,y index=1]{final_success_ratio.csv};

\addplot[thin, green!60!black]
table[col sep=comma,x index=0,y index=3]{final_success_ratio.csv};

\addplot[thin, purple]
table[col sep=comma,x index=0,y index=2]{final_success_ratio.csv};

\addplot[thin, orange]
table[col sep=comma,x index=0,y index=4]{final_success_ratio.csv};

\end{axis}
\end{tikzpicture}

\centerline{\scriptsize (a) Success vs.\ time}
\end{minipage}

&

\refstepcounter{subfig}\label{fig:avg-hop}
\begin{minipage}{0.48\columnwidth}

\centering
\begin{tikzpicture}
\begin{axis}[
    width=4.6cm,
    height=3.6cm,
    xlabel={Payment Size (sat)},
    ylabel={Avg Hops},
    ymin=2.5,
    ymax=4.5,
   xtick={1,2,3,4,5},
        xticklabels={
        \shortstack{$10^0$--\\$10^2$},
        \shortstack{$10^2$--\\$10^3$},
        \shortstack{$10^3$--\\$10^4$},
        \shortstack{$10^4$--\\$10^5$},
        \shortstack{$10^5$--\\$10^6$}
        },
        xticklabel style={
            font=\scriptsize,
            align=center
        },
        xticklabel style={font=\tiny, align=center},
    grid=both,
    grid style={gray!20},
    tick label style={font=\scriptsize},
    label style={font=\scriptsize},
    legend style={font=\scriptsize, at={(0.5,1.12)}, anchor=south, draw=none},
    clip=false
]
\addplot[thick, violet, mark=triangle*]
table[x expr=\coordindex+1,y index=1,col sep=comma]{payment_vs_hops.csv};

\addplot[thick, yellow, mark=diamond*]
table[x expr=\coordindex+1,y index=1,col sep=comma]{payment_vs_hops_2.csv};

\addplot[thick, blue, densely dashed, mark=square*]
table[x expr=\coordindex+1,y index=1,col sep=comma]{payment_vs_hops_coalesce.csv};

\end{axis}
\end{tikzpicture}

\vspace{-0.3em}
\centerline{\scriptsize (b) Avg hops vs.\ payment size}
\end{minipage}

\\

\refstepcounter{subfig}\label{fig:depletion_vs_iter}
\hspace{1.1em}
\begin{minipage}{0.48\columnwidth}
\centering
\vspace{-0.5em}
\begin{tikzpicture}
\begin{axis}[
    width=4.6cm,
    height=3.6cm,
    xlabel={Iteration},
    ylabel={Depletion},
    xmin=1, xmax=200,
    grid=both,
    grid style={gray!20},
    tick label style={font=\scriptsize},
    label style={font=\scriptsize},
    legend style={font=\scriptsize, at={(0.5,1.22)}, anchor=south, draw=none},
    clip=false,
    scaled y ticks=base 10:-3,
    y tick label style={
        /pgf/number format/fixed,
        /pgf/number format/precision=2
    }
]

\addplot[semithick, blue, densely dashed]
table[x expr=\coordindex+1, y index=1, col sep=comma]{depletion_combined.csv};

\addplot[thin, purple]
table[x expr=\coordindex+1, y index=3, col sep=comma]{depletion_combined.csv};

\addplot[thin, green!60!black]
table[x expr=\coordindex+1, y index=2, col sep=comma]{depletion_combined.csv};

\end{axis}
\end{tikzpicture}

\centerline{\scriptsize (c) Depletion vs.\ iteration}
\end{minipage}

&

\refstepcounter{subfig}\label{fig:success_vs_lrf}
\begin{minipage}{0.48\columnwidth}
\centering

\hspace{-1.3em}
\begin{tikzpicture}
\begin{axis}[
    width=4.6cm,
    height=3.6cm,
    xlabel={LRF},
    ylabel={Success (\%)},
    grid=both,
    grid style={gray!20},
    tick label style={font=\scriptsize},
    label style={font=\scriptsize},
    legend style={font=\scriptsize, at={(0.5,1.12)}, anchor=south, draw=none},
    clip=false
]
\addplot[blue, thin, smooth, densely dashed]
table[col sep=comma,x index=1,y index=0]{success_vs_lrf.csv};
\end{axis}
\end{tikzpicture}

\centerline{\scriptsize (d) Success vs.\ LRF (long-term run)}
\end{minipage}

\end{tabular}
}
\caption{Full hypergraph evaluation.}
\label{fig:hypergraph-eval}
\end{figure}


The success-ratio comparison (Fig.~\ref{fig:success_vs_time}) shows that our model sustains consistently high reliability across iterations.  
Since the protocol does not rely on channel rebalancing or virtual channel overlays, Horcrux appears as a natural baseline for comparison, and \prot\ achieves comparable or stronger stability without additional coordination mechanisms.

The average-hop analysis (Fig.~\ref{fig:avg-hop}) directly compares \prot\ with Ticket To Ride and Speedy Murmurs.  
We use empirical values reported in the Ticket To Ride paper for the baselines.  
\prot\ achieves an average hop count in the range $2.7$--$2.5$, which is strictly lower than both baselines (lower is better), indicating significantly stronger effective connectivity than traditional PCN-style routing.

The depletion analysis (Fig.~\ref{fig:depletion_vs_iter}) compares against Shaduf and OPT-Revive.  
Although \prot\ does not use channels, each common participant effectively plays the role of a bidirectional liquidity carrier between hyperedges.  
Since multiple intermediaries can coexist between hyperedges, liquidity flow is distributed rather than concentrated, resulting in substantially lower effective depletion over time.

Finally, the long-term experiment (100,000 transactions) relates success ratio to Liquidity Retention Factor (Fig.~\ref{fig:success_vs_lrf}).  
Despite sustained heavy load, the system maintains success rates around $84\%$ even as some intermediaries experience partial liquidity erosion, indicating robustness under prolonged operation.

Taken together, these results suggest that the protocol scales from single-hyperedge execution to full multi-hyperedge deployment without introducing the structural fragility observed in channel-based payment networks.

%% file: paper/sections/related_work.tex
\section{Related Work}
\label{sec:related}

Payment Channel Networks (PCNs) were introduced to improve blockchain scalability by moving transactions off-chain while preserving on-chain security guarantees. Early systems such as Lightning~\cite{poon2016lightning}, Sprites~\cite{Sprites}, and Raiden~\cite{raidennetwork} established the basic model of multi-hop payments using hashed timelock contracts. Subsequent work focused on routing efficiency and reliability, including Spider~\cite{SpiderRouting}, Boomerang~\cite{Boomerang}, Flare~\cite{Flare}, and Ticket to Ride~\cite{t2r2017}. A large body of work studied liquidity imbalance and channel exhaustion, proposing rebalancing or virtual channel mechanisms such as Revive~\cite{Revive}, REBAL~\cite{rebal3}, Shaduf~\cite{10606068}, Perun~\cite{8835315}, Thora~\cite{DBLP:journals/iacr/AumayrAM22}, and Bitcoin-compatible virtual channels~\cite{aumayr2021bitcoinvc}. While effective in specific settings, these approaches typically rely on explicit channel structures, coordination, or additional protocol layers that limit scalability under sustained load.

More recently, hypergraph-based payment networks have been proposed to generalize channel connectivity and improve scalability. Kotzer et al.~\cite{hpn} study hypergraph payment networks topology and analyze their potential to reduce routing bottlenecks by allowing multi-party connectivity. Our work differs fundamentally in design: \prot\ realizes hyperedge semantics through proposer-ordered DAGs with threshold signed DAG roots, and supports inter-hyperedge transfers using locally verifiable PoTs (\emph{proofs-of-transfer})  rather than global coordination or channel reconfiguration. This allows inter-hyperedge transactions while ensuring liveness and consistency.

%% file: paper/sections/future_work.tex
\section{Future Work}
\label{sec:future}
Future work includes a detailed evaluation of protocol performance under realistic network conditions and varying hyperedge sizes, as well as refinement of incentive mechanisms for participation and inter-hyperedge relaying. Implementing the protocol using Taproot-based primitives would enable measurement of on-chain dispute costs and practical deployment considerations. Another direction is the study of liquidity allocation and topology formation in networks composed of overlapping hyperedges, which may influence payment success rates and routing flexibility.

%% file: paper/sections/conclusion.tex
\section{Conclusion}
\label{sec:conclusion}
This work introduced \prot, a protocol that represents jointly funded off-chain channels as hyperedges and maintains state using proposer-chained DAG nodes with revocation-based ordering. The protocol enables consistent intra-hyperedge payments and supports inter-hyperedge transfers through common participants using PoTs and finalized DAG root state transitions. Threshold-signed DAG roots provide a verifiable settlement mechanism, while DAG nodes which have revocation secrets allow strong dispute evidence in case of misbehavior. Together, these components yield a decentralized and auditable off-chain value transfer protocol that preserves safety under standard Byzantine assumptions.